# Probing exciton localization in non-polar GaN/AlN Quantum Dots by single dot optical spectroscopy

F. Rol[1], S. Founta[1,2], H. Mariette[1,2], B. Daudin[1], Le Si Dang[2], J. Bleuse[1], D. Peyrade[3], J.-M. Gérard[1] and B. Gayral[1,*]

[1]Equipe CEA-CNRS-UJF Nanophysique et Semiconducteurs, DRFMC/SP2M/PSC, CEA-Grenoble,17 Rue des Martyrs, 38054 Grenoble Cedex 9, France

[2]Equipe CEA-CNRS-UJF Nanophysique et Semiconducteurs, Laboratoire de Spectrométrie Physique (CNRS UMR 5588), Université Joseph Fourier, 38402 Saint Martin d'Hères, France

[3]Laboratoire des Technologies de la Microélectronique (LTM/CNRS), 17 Avenue des Martyrs (CEA-LETI), 38054 Grenoble Cedex 9, France

**Abstract**

We present an optical spectroscopy study of non-polar GaN/AlN quantum dots by time-resolved photoluminescence and by microphotoluminescence. Isolated quantum dots exhibit sharp emission lines, with linewidths in the 0.5-2 meV range due to spectral diffusion. Such linewidths are narrow enough to probe the inelastic coupling of acoustic phonons to confined carriers as a function of temperature. This study indicates that the carriers are laterally localized on a scale that is much smaller than the quantum dot size. This conclusion is further confirmed by the analysis of the decay time of the luminescence.

## I- Introduction

III-N heterostructures are usually grown along the polar [0001] axis (c-axis), giving rise to a built-in internal field in quantum wells (QWs) or quantum dots (QDs) [1] due to the polarization discontinuity along the c-axis at the well-barrier interface. For quantum wells grown along a non polar axis, the c-axis, being in the plane of the well, does not cross any polarization discontinuity so that such a quantum well is free from internal electric field [2]. For III-N quantum dots grown along a non polar axis, the situation is not as straightforward as the c-axis still crosses a polarization discontinuity at the quantum dot lateral facets. The build-up of an internal field is then very sensitive to the exact dot geometry (shape and strain)

---
[*] Corresponding author. Electronic address : bruno.gayral@cea.fr

[3] as well as to possible screening due to residual doping. In this article, we study GaN in AlN quantum dots grown along the [11-20] axis (a-axis). While a strongly reduced quantum confined Stark effect (QCSE) compared to polar QDs of similar sizes was already observed [4, 5], the optical study presented here aims at probing the lateral extent of the confined wavefunctions using time-resolved photoluminescence and single QD photoluminescence spectroscopy.

## II- Samples and experiments
### A- Samples and optical properties of QD ensembles

The sample used for the present study was grown by plasma assisted molecular beam epitaxy (PA-MBE), on a (11-20) 6H-SiC substrate provided by INTRINSIC Semiconductor and polished by NOVASiC. It consists of a 30 nm AlN buffer layer followed by a first array of self organized GaN quantum dots, then a 20 nm AlN cap layer is grown followed by a surface quantum dots array similar to the first one. The 2D-3D transition was controlled during the growth by the appearance of three dimensional RHEED (reflection high-energy electron diffraction) patterns, which happens after the deposition of 3.7 monolayers of GaN. Details of the growth procedure can be found elsewhere [4]. The uncapped quantum dots, studied by atomic force microscope (AFM) show a rather high areal density of $1.2 \, 10^{11}$ cm$^{-2}$, a typical height of 2 nm and a diameter of 20 nm [4].

The samples were mounted in a helium-flow cryostat for time-resolved and time-integrated photoluminescence (TRPL and PL) measurements and the temperature could be controlled from 5 K to more than 300 K. We present in Fig. 1 time-resolved and time-integrated photoluminescence measurements of the buried quantum dots plane at 5 K. Experiments were performed using a tripled Ti-sapphire laser of average power 3 mW at the wavelength of 250 nm and pulsed at the frequency of 76 MHz. The quantum dots luminescence was analyzed using a 320 mm monochromator and a streak camera with a time resolution of 5 ps.

Compared to the optical properties of the polar GaN quantum dots (0001), the non-polar ones (11-20) present three main features evidencing a much weaker QCSE, as already observed in previous works [4,5]. First, the PL intensity presents a maximum around 3.8 eV and even for large dots, the emission is at higher energy than the GaN strained bandgap. This clearly shows that quantum confinement effects are dominant over any quantum confined Stark effect, by contrast to (0001) QDs of similar dimensions [6, 7]. Secondly, no blue shift has been reported under high excitation power [5], contrary to the polar dots for which the internal field can be screened by the photocreated carriers [7, 8]. And thirdly, previous studies [4] show that non-polar GaN quantum dots present PL decay times at low temperature in the 200-500 ps range instead of a few ns and more for the (0001) dots [6, 9]. Indeed, the presence of an internal electric field for c-plane QDs results in a spatial separation of the electron and hole wave functions, thus decreasing the oscillator strength of the transitions.

For the samples studied in this article, the decay curves do not depend on the emission wavelength (Fig. 1). This is again in strong contrast with polar QDs, for which the quantum confined Stark effect depends on the QD size : there is thus a strong dependency of the decay rate as a function of the emission wavelength [9]. The decay curves we observe are not mono-exponential : the decay on the short time-scale is about 70 ps while the decay on the longer time-scale is about 600 ps, the mean decay time being 280 ps. We shall discuss these decay values later in this article. The decay curves remain stable up to 100 K, so that we can exclude the role of non radiative processes in the low temperature recombination dynamics. As a first conclusion, the luminescence decay in the studied a-plane samples at low temperature is radiative, with much shorter decay values than for c-plane QDs, the radiative lifetimes being independent of QD size. The optical characterization of ensembles of a-plane QDs and especially the time-resolved experiments do not show any evidence of QCSE.

### B- Microphotoluminescence properties

In order to access the optical properties of a single GaN quantum dot, we aim at isolating the luminescence of the smallest possible number of dots by etching mesas [10] or by opening small apertures in an aluminum mask [11]. For the mesas, square openings in a resist layer with sizes ranging from 5 μm down to 200 nm were designed by E-beam lithography. After deposition of a 50 nm thick nickel film and a lift-off, the mesas were formed by $SiCl_4$ inductively coupled plasma (ICP) etching. The nickel mask was then removed in a $HNO_3$ solution. For the aluminum mask, we put a drop of polystyrene nano-beads (diameter=300 nm) dispersed in water on the sample and let it dry, then we deposited 100 nm aluminum and subsequently removed the beads to leave apertures in the mask [10]

As the average density of dots is rather high ($5\ 10^{10}$ to $1.2\ 10^{11} cm^{-2}$) the smallest holes or mesas contain on average 30 to 80 dots. However, it is possible to find mesas and apertures containing only a few dots since the areal density of dots is quite inhomogeneous on a ~1 μm length scale for this sample.

Single QDs are studied by low temperature microphotoluminescence. The excitation is provided by a cw doubled argon laser line at 244 nm creating electron-hole pairs in the wetting layer. The beam is focused on the sample by a UV-optimized microscope objective (NA=0.4) leading to a 1 μm spot. The photoluminescence is collected through the same microscope objective, analyzed by a single grating monochromator (resolution 0.3 meV around 4 eV) and detected by a liquid nitrogen cooled CCD camera.

The photoluminescence of the smallest mesas presents isolated sharp lines in the high energy tail of the quantum dot distribution (>3.8 eV), but for reasons that are still not fully understood we could not find such sharp structures on the low energy side of the spectrum. We should mention that these sharp lines display a linear dependence with the excitation laser at low excitation power, so that they can be attributed to fundamental single electron-hole pair transitions of the QDs. In general, the linewidths of the studied isolated peaks are between 0.5 and 2 meV, which is much larger than the lifetime limited linewidth (5 µeV). We attribute this broadening to a Coulomb interaction between the confined exciton and loosely trapped charges moving in the vicinity of the dot leading to a time-dependent spectral diffusion [13, 14]. All phenomena that occur on a shorter time-scale than our accumulation time (typically 1s) are averaged on the spectra. Even when accumulating several spectra one after the other with integration time of 1 s, a spectral diffusion of the lines of the order of 1 meV can be observed (Fig. 2). Although larger than for III-As systems, such linewidths are one order of magnitude smaller than values reported for single polar GaN QDs [15]. This is consistent with the fact that the internal electric field for c-plane QDs should enhance the effect of the electrostatic environment fluctuations on the quantum dot transition and thus amplify the spectral diffusion.

The lines obtained in these non-polar GaN single quantum dots are narrow enough to be sensitive to temperature induced broadening mechanisms. In Fig. 3 we present a single dot spectrum taken at different temperatures from 5 K up to 140 K and at low excitation power. The spectra are normalized to the spectrally integrated intensity. Below 60 K, the line shape does not evolve: spectral diffusion is the dominant broadening mechanism. Above 60 K, temperature induced broadening takes over as the dominant broadening mechanism. This includes the appearance of asymmetric phonon wings (Fig. 4) due to absorption and emission of acoustic phonons in the luminescence process (inelastic phonon scattering). As detailed in section III, the analysis and modeling of these phonon wings allow to finely probe the confined electronic wavefunctions.

**III-Modeling and analysis : Localization of the exciton**
**A-    Exciton-phonon interaction**

The observation of phonon wings on single QD spectra when raising the temperature has already been reported for CdTe [16], InAs [17, 18] and GaAs [19] QDs.

This mechanism was first described and modeled by Besombes and coworkers [16] by extending the Huang-Rhys theory of localized electron-phonon interaction to the case of an exciton localized in a quantum dot. Let us remind the framework of this model that we shall apply to our particular experimental situation. In this model, the dispersion of the acoustic phonons is discretized into N modes, and for each of them the Huang-Rhys theory defines new eigenstates where the exciton and each monochromatic acoustic phonon mode are in strong coupling: the acoustic polaron states. This nonperturbative coupling creates a discrete set of polaron states which can recombine radiatively but with different probabilities depending on the phonon part of each exciton-acoustic-phonon state. The exciton-phonon interaction Hamiltonian is given by: $H_{X-LA-ph} = c^+ c \sum_{\vec{q}} M_{\vec{q}} (b_{\vec{q}}^+ + b_{\vec{q}})$ where $c^+$ and $b_q^+$ ($c$ and $b_q$) are respectively the creation (annihilation) operators of the ground state exciton $|X\rangle$ and of the phonon (of wave vector q and energy $\hbar\omega_q$). To calculate the matrix element $M_q$, only the deformation potential induced by the longitudinal acoustic phonon is considered. This approximation is in general valid for zinc-blende compounds for which the piezo-electric terms are much smaller than the deformation potential terms [20, 21]. For bulk III-N compounds, the piezo-electric coupling parameters are much larger so that neglecting these terms is not obvious. While the deformation potential coupling is mainly sensitive to the overall spatial extent of the confined electron and hole wavefunctions, the piezo-electric coupling is sensitive to the wavefunction difference between the hole and the electron (in case of perfect local neutrality, the piezo-electric coupling vanishes) [20, 21]. In particular in the case of an electron-hole spatial separation due to a static electric field, the piezo-electric coupling is much enhanced [21, 22]. In the case of the non-polar QDs discussed in this article, as there is no effect of the internal electric field in the cw and time-resolved characterization on ensembles of dots, it can be assumed that the permanent dipole of the confined electron-hole pair is small enough so that the piezo-electric coupling to the acoustic phonons can be neglected compared to the deformation potential coupling. This assumption is justified *a posteriori* by the good fit obtained for the temperature dependent phonon wings, and by the consistency of the global picture we propose of an electron-hole pair exhibiting a strong lateral confinement and a negligible internal electric field effect.

It is also important to remind that the following results are valid as long as the separation between the ground and the excited states are large compared to the acoustic phonon energies as no mixing between electronic states is taken into account. $M_q$ is thus given by: $M_{\vec{q}} = \sqrt{\frac{\hbar|\vec{q}|}{2\rho u_s v}} \left( D_c \langle X | e^{i\vec{q}\cdot\vec{r}_e} | X \rangle - D_v \langle X | e^{i\vec{q}\cdot\vec{r}_h} | X \rangle \right)$, where $\rho$ is the mass density, $u_s$ the isotropic averaged sound velocity stemming from a Debye approximation for the LA-phonon relation dispersion ($\omega_q = u_s q$), v the quantization volume, and $D_c$ ($D_v$) the deformation potential of the conduction (valence) band.

As the Bohr radius of the GaN bulk exciton is $a_{GaN}=2.8$ nm [23], the exciton located in our quantum dots is still strongly confined along the height (~2 nm), but weakly confined in the a-plane (QD diameter~20 nm). As a result the exciton state $|X\rangle$ should *a priori* be well described by a quasi-two-dimensional wave function, laterally limited by a Gaussian distribution of the center of mass.

$$\psi_X(r_e, r_h) = \left( \frac{1}{\xi\sqrt{\pi}} e^{-R_p^2/2\xi^2} \right) \left( \sqrt{\frac{2}{\pi a_{2D}^2}} e^{-r/a_{2D}} \right) \left( \frac{2}{L} \cos(\pi z_e / L_Z) \cos(\pi z_h / L_Z) \right) \quad (1)$$

The third part of this wave function is the solution of an infinite barrier quantum well with a height of $L_z$, which is a good approximation because band offsets between GaN and AlN are large enough for AlN to behave as an infinite barrier. The in-plane electron-hole correlation is described by a 2-D hydrogenoid wave function, **r** being the electron/hole relative position and **a$_{2D}$** the 2D Bohr radius of the infinite quantum well. The lateral extension of the center of mass position **R$_p$** is described by the localization parameter $\xi$.

With these assumptions, $M_q$ can be calculated, $\xi$ being the only fitting parameter of the model. The most relevant parameter describing the exciton-phonon coupling is the Huang-Rhys factor g(q) obtained by integrating over all directions of **q** the coupling constant defined by: $g_{\vec{q}} = |M_{\vec{q}}|^2 / (\hbar\omega_{\vec{q}})^2$. The study of g(q) shows that the coupling is maximum for phonons with $q \sim 1/\xi$ and becomes negligible for $q > 2/\xi$. The maximum value of g(q) increases for more localized wavefunctions (smaller $\xi$).

As shown first by Huang and Rhys [24], the probability $W_p^q$ that the optical transition involves p phonons with a wavevector modulus q can be calculated from g(q). The expression for $W_p^q$ can be found in Ref. 16. p can be positive or negative, corresponding respectively to the emission or the absorption of p phonons. $W_0^q$ is the probability to recombine without phonon emission or absorption (zero phonon line). To apply the model to LA acoustic phonons, we discretize the dispersion relation into N=50 modes in the wavevector regions where the phonons couple efficiently with the confined exciton. It is then possible to calculate the spectral shape function by the combination of every simultaneous emission probability $\prod_{1<i<50} W_{p_j}^{q_i}$ for every set of $p_j$. We however checked that in the temperature range that we consider here (< 200K) the probability of a transition involving more than 3 phonons is negligible so that such events need not be taken into account in the calculation. The calculated spectral shape function is a set of $\delta$-like functions representing each the probability to recombine with a combination of different LA-phonons. To finally get an emission spectrum simulation, we convolute the spectral shape function by the zero-phonon line.

As the spectral diffusion is important in our GaN QDs, the zero-phonon line has no reason to be well described by a simple Lorentzian. Indeed at low temperature for which the spectral diffusion is the dominant broadening process, the line-shape is mainly gaussian with a width $w_g$=1.1 meV. To fit the zero-phonon line we thus use a Voigt function with FWHM $w_g$ and $w_l$ for the Gaussian and the Lorentzian part respectively. $w_g$ corresponds to the spectral diffusion, and $w_l$ is related to the additional temperature dependent broadening mechanisms of the zero-phonon line.

For the calculations, we used the parameter values $D_c$=-9 eV, $D_v$=1 eV [25], $\rho$=6150 g.cm$^{-3}$, $u_s$=8000 m.s$^{-1}$, $m_e$=0.2$m_0$, $m_h$=$m_0$. For a luminescence energy of E=3.98 eV a simple infinite barrier quantum well model gives an estimated height $L_z$=1.6 nm, and a value of the two dimensional Bohr radius $a_{2D}$=2.1 nm. We set $w_g$ to the low temperature experimental value of 1.1 meV for every temperature, assuming that the spectral diffusion is constant with temperature. Now the way to describe the temperature dependent behavior of the single exciton line with this model consists in finding the correct localization parameter $\xi$, and to change only $w_l$ at each temperature to take the zero-phonon line broadening into account.

By following this procedure we find a localization parameter $\xi$=2.1 nm, and an increase of $w_l$ from 0.7 meV at 5 K to 2.4 meV at 140 K. The increase of the Voigt function linewidth ($\sqrt{w_l^2 + w_g^2}$) with temperature is larger than expected in a quantum dot, but comparable to already reported results in Ref. [26]. Fig. 4 presents the experimental points at three different temperatures compared with the calculated spectra. In order to visualize the amplitude and the asymmetry of the phonon sidebands, Fig. 4 also represents the zero-phonon line in a dotted line. The theoretical curves fit very well the experimental data over the experimentally accessible temperature range.

The probing of the exciton wavefunction in a-plane QDs by analysis of the coupling to acoustic phonons leads thus to the determination of the lateral localization parameter $\xi$=2.1 nm, which corresponds to a full width at half maximum of the center of mass wavefunction of 4.9 nm. The exciton is thus localized on a lateral scale which is much smaller than the average QD diameter of 20 nm. Further evidence of this localization and discussion of possible localization mechanisms are presented below.

### B- Decay time analysis

Another way of probing the exciton confinement in a "flat" QD is to analyze its radiative decay time. Two extreme confinement regimes are of particular interest. In the "strong confinement regime", the electron and the hole are localized on a scale that is smaller than the Bohr radius of the QW exciton. In that case, excitonic effects become negligible. The exciton oscillator strength (or equivalently its radiative lifetime) does not depend explicitly on the QD size, but only on the electron-hole wavefunction overlap. In the opposite case, for which the exciton is loosely confined laterally (with respect to the scale of the Bohr radius of the QW exciton), the oscillator strength increases proportionally to the coherence volume of the exciton. This regime which has been studied for flat QDs by Kavokin [27] and Andreani and coworkers [28], is known as the "giant oscillator strength regime". Following the approach developed in Refs. [27, 28], we find that for the quasi two-dimensional wavefunction (1) and for $2\xi > a_{2D}$, the oscillator strength is given by:

$$f = \frac{8E_p}{E} \cdot \left(\frac{\xi}{a_{2D}}\right)^2 \quad (2)$$

where E is the exciton energy, and $E_p$ the interband matrix element (Kane energy). The Kane energy for the "heavy hole" valence band in wurtzite GaN was calculated by Chang and Chuang to be 15.7 eV [29].

In order to discuss the nature of the exciton confinement in our GaN QDs, let us now calculate the oscillator strength and radiative exciton lifetime for three different cases : i) an exciton wavefunction that extends laterally over the entire quantum dots, its coherence volume being only limited by the lateral confining potential of the AlN barrier ii) an exciton wavefunction with an additional lateral localization mechanism as deduced from the single dot phonon coupling experiments iii) a laterally strongly confined exciton for which the confinement effects dominate over the Coulomb effects. Let us first recall at this point that the radiative lifetime $\tau_r$ is related to the oscillator strength f by :

$$\tau_r = \frac{3m_e c^3 4\pi\varepsilon_0}{2ne^2\omega^2 f} \quad (3)$$

where $m_e$ is the electron mass and n the refractive index of the medium.

In the first case i), we expect the QD to be in the giant oscillator strength regime, since its lateral size is much larger than the Bohr radius of the QW exciton. More precisely, the transmission electron microscopy (TEM) and high resolution AFM images show QDs with truncated pyramid shapes (rectangular base), with facets orientated at about 30 degrees. Given the strong bandgap difference between AlN and GaN, the center of mass of the exciton can be considered to be laterally limited by an infinite barrier in a cylinder of diameter corresponding to the top dimension of the truncated pyramid. The top diameter $D_t$ of the QD is given by the geometric relation $D_t = D - 2.\sqrt{3}.h$ where D and h are respectively the base diameter and the height of the QD. In this simple model, the localization parameter $\xi$ is given by the gaussian the closest to the sinusoidal shape of the center of mass exciton wavefunction in the infinite barrier potential. This approximation underestimates $\xi$ but leads approximately to $\xi \approx D_t / 4$. By applying this formula to the center of the QD distribution: h=1.6 nm and D=19 nm, we obtain $\xi \approx 3.4$ nm. From equations (2) and (3) we deduce $f \approx 85$ and thus $\tau_r \approx 24$ ps. This estimate of $\tau_r$ is well below experimentally measured values although our wavefunction calculation underestimates $\xi$ and thus overestimates $\tau_r$. This is consistent with the fact that an additional exciton localization mechanism occurs in these QDs.

In case iii), the oscillator strength tends towards the value $f = \frac{E_p}{E}$ [28] for the ideal case of a perfect electron-hole overlap. This gives in our case an oscillator strength of 4 or conversely a decay time value of $\tau_r \approx 520$ ps, which is in the upper decay time values that we experimentally measure for an ensemble of QDs. The actual situation is thus intermediate between the strong confinement regime and the quasi-2D case.

In case ii), with the localization parameter found in the µ-PL studies (ξ=2.1 nm), we deduce from equations (2) and (3) an oscillator strength f ≈ 33 and a decay time $\tau_r$ ≈ 62 ps. This value is still a factor of 4 smaller than the measured average decay time, which calls for a comment. Firstly, the Kane energy might be smaller than the value we used, while the authors of Ref. 29 calculate a value of 15.7 eV and the authors of Ref. 25 recommend a similar value of 14 eV, another study deduced a value of 7.7 eV [30]. For this lower value of $E_p$, the measured oscillator strength would still be between the strong confinement limit and the quasi-2D framework with ξ=2.1 nm, but less towards the strong confinement limit. Secondly, the strong enhancement factor that appears in formula (2) is due to the excitonic nature of the quasi-2D wavefunction. It is clear that when ξ is comparable with the 2D Bohr radius, the wavefunction cannot anymore be factorized into an excitonic part and a center of mass motion, so that the excitonic enhancement of the oscillator strength is reduced.

At this point we recall that in the deformation potential coupling of acoustic phonons to the localized carriers, the excitonic nature of the electron-hole pair does not play an important role. Through the deformation potential the acoustic phonons are sensitive mainly to the overall extent of the localized electron and hole wavefunctions so that the subtleties of the interplay between Coulomb and confinement energies do not affect much the model.

Finally, let us recall that the measured decay curve for an ensemble of QDs at a given emission wavelength is not mono-exponential and is better described by a distribution of decay times ranging from 70 to 600 ps. We then probed the photoluminescence decay on single QDs (Fig. 5) by adapting the microphotoluminescence set-up on the time-resolved experiment. These experiments are difficult to perform due to the low detection sensitivity of the streak camera. Nonetheless, sharp peaks could be isolated and studied as a function of time. Each single peak decays monoexponentially, but the decay times vary from peak to peak, which confirms that each peak can be attributed to the emission of a single QD. The decay time for each QD is not correlated to its emission wavelength. These observations suggest that excitons are further randomly localized by an additional potential in each individual QD. Thus the QD exciton lifetime is not governed by the lateral extension of the QD, but by the lateral extension of its wavefunction resulting from this additional localization mechanism.

Let us sum-up the key results and conclusion so far exposed in this article. i) In microphotoluminescence, the study of the coupling of a confined electron-hole pair to acoustic phonons is consistent with a quasi-two dimensional wavefunction with a localization parameter ξ=2.1 nm. This could however only be shown for the smallest QDs emitting in the high energy part of the QD emission spectrum. ii) The analysis of the oscillator strength confirms that for all dot sizes the wavefunction is more laterally localized than expected from the sole AlN barrier lateral confinement.

We shall now compare these conclusions with results obtained on other III-N heterostructures. For instance a-plane GaN/AlGaN quantum wells were studied in Ref. 31, in particular by using time-resolved spectroscopy. The authors of Ref. 28 discuss the variation of decay time as a function of QW width but do not discuss the absolute value of the decay time. Indeed, for the thinnest QW in Ref. 31 that has a height comparable to our QD samples, the decay time at low temperature is similar (within the experimental accuracy) to the one we find for QDs. This corresponds to a long decay time for a QW emission, which would be consistent with a lateral localization on dimensions comparable with the Bohr radius. We also note that the lifetime measured in cubic phase GaN/AlN QDs, for which there is no internal field, is about 200 ps [6] and thus similar to the decay times measured here. Again the giant oscillator strength regime is not reached for these cubic QDs although they are quite large (3 nm height, 30 nm wide) so that an additionnal localization should occur. We also studied a 1.5 nm thick GaN/AlN QW grown along the a-axis. The decay curve for the photoluminescence at 5K is presented in the inset of Fig. 6 and compared to the decay curve for a-plane QDs. The decay curve for the QW is not monoexponential. Still an average decay time can be extracted and is reported as a function of temperature in Fig. 6. The first striking feature is that the photoluminescence decay curves do not change between 5 K and 100 K, which indicates that the excitons are laterally localized. The low temperature decay time (300 ps) corresponds to a small oscillator strength for a QW emission, again leading to the conclusion that the QW exciton has a small coherence surface and that a strong lateral localization occurs. We can deduce that in our a-plane samples, at low temperature the excitons are laterally localized in QWs as in QDs. This is further confirmed by the fact that the decay curves remain the same between 5 K and 100 K for both the QW and the QDs emissions. Above 100 K, delocalization probably occurs, leading to a stronger non-radiative decay probability for the QW than for the QDs.

**Conclusion**

All these results are consistent with the hypothesis that a strong lateral localization comparable with the two-dimensional Bohr radius occurs in GaN/AlN heterostructures for state of the art MBE growth. Such localization could for instance be induced by interface fluctuations. Localization effects were already pointed out for instance in shallow polar GaN/AlGaN quantum wells [32], however in that case alloy fluctuations in the ternary barrier alloy are likely to be responsible for localizing potentials. If this localizing effect also occurs in polar GaN/AlN heterostructures, then for the larger QDs the electrons and holes localized at the top and bottom of the QDs by the internal electric field, might be laterally separated as well. This effect might have an important role on the very low oscillator strengths reported for large polar GaN/AlN QDs [9].

**Acknowledgements:**

We acknowledge Marlène Terrier and Joël Eymery for experimental contribution. This work is supported by the ACI "BUGATI" and by Université Joseph Fourier (Grenoble) through the "BIGAN" project.

**Figure 1 :**

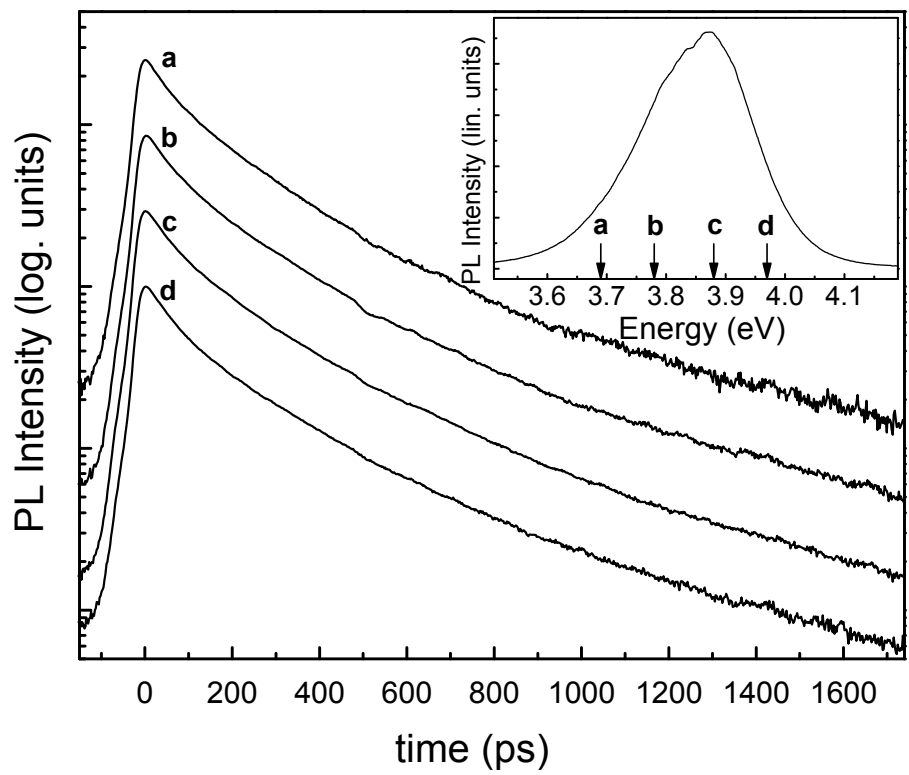

**Figure 2:**

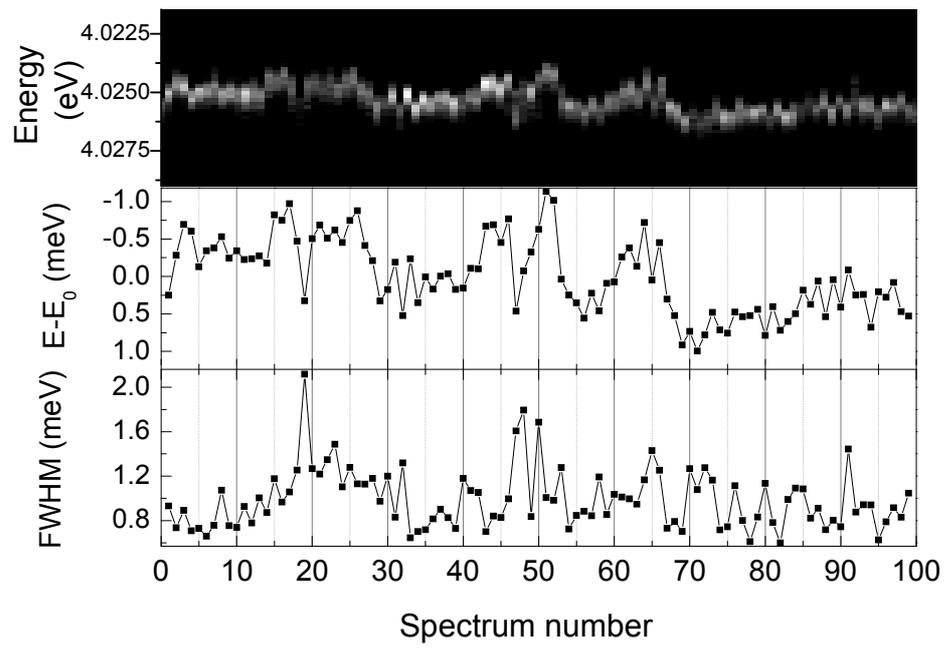

**Figure 3 :**

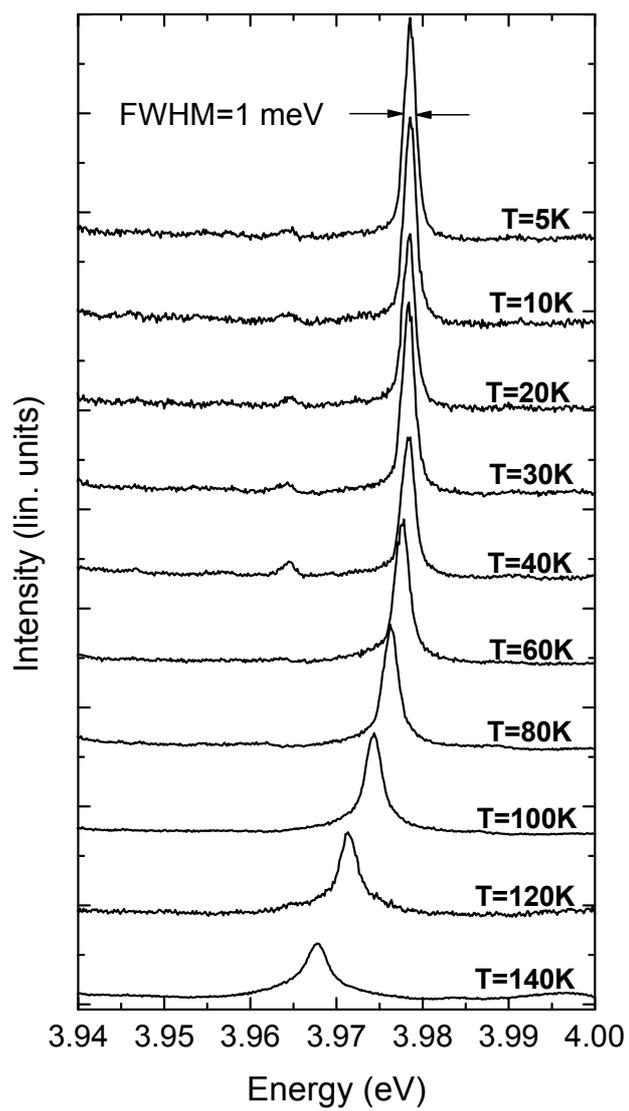

**Figure 4 :**

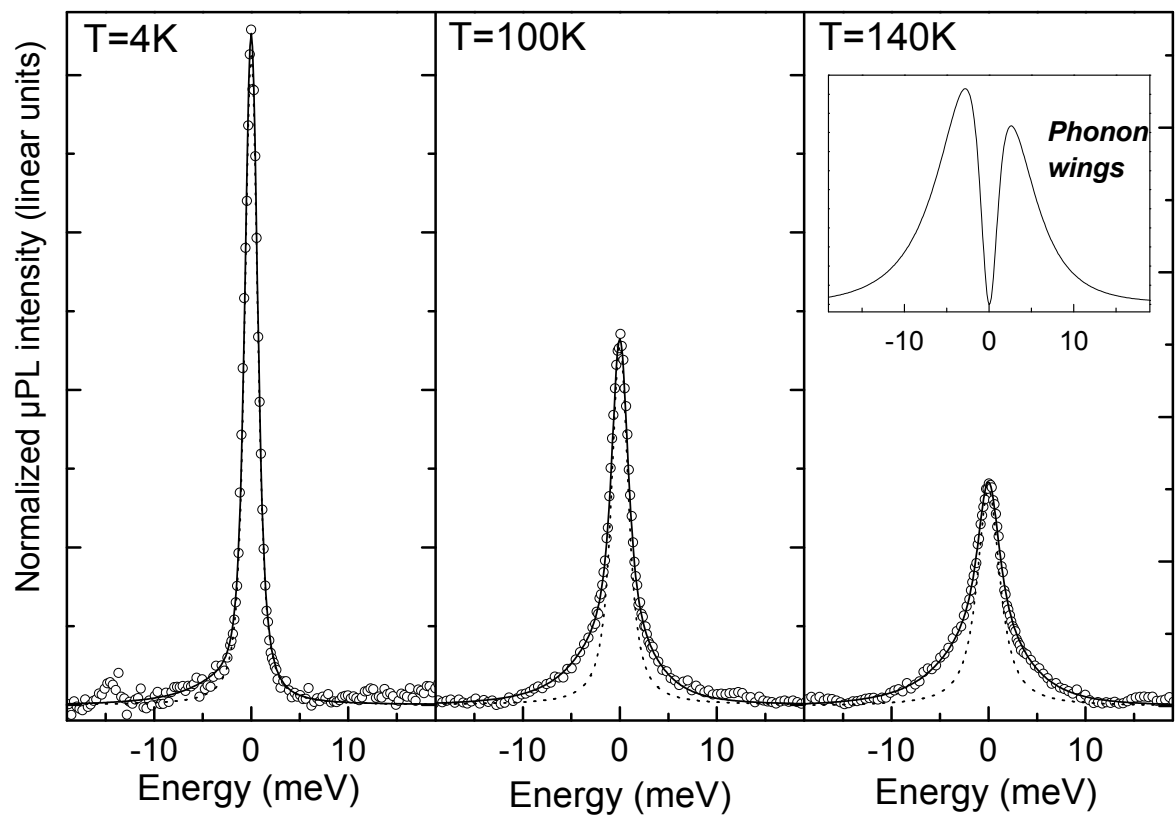

**Figure 5 :**

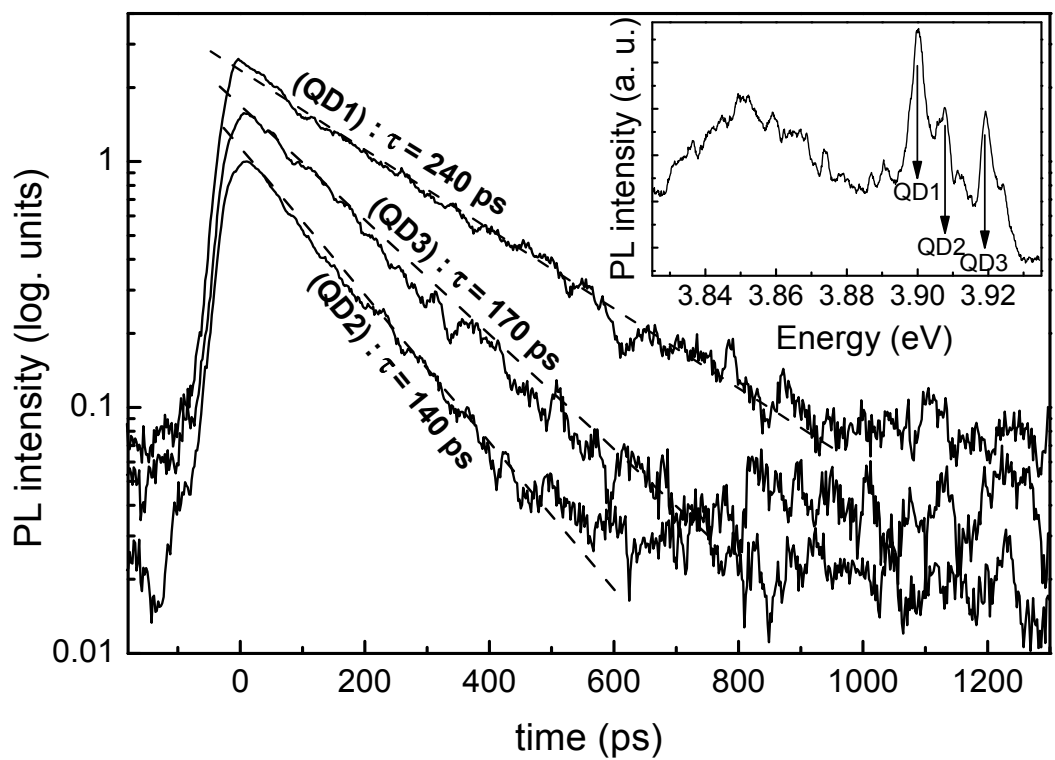

**Figure 6 :**

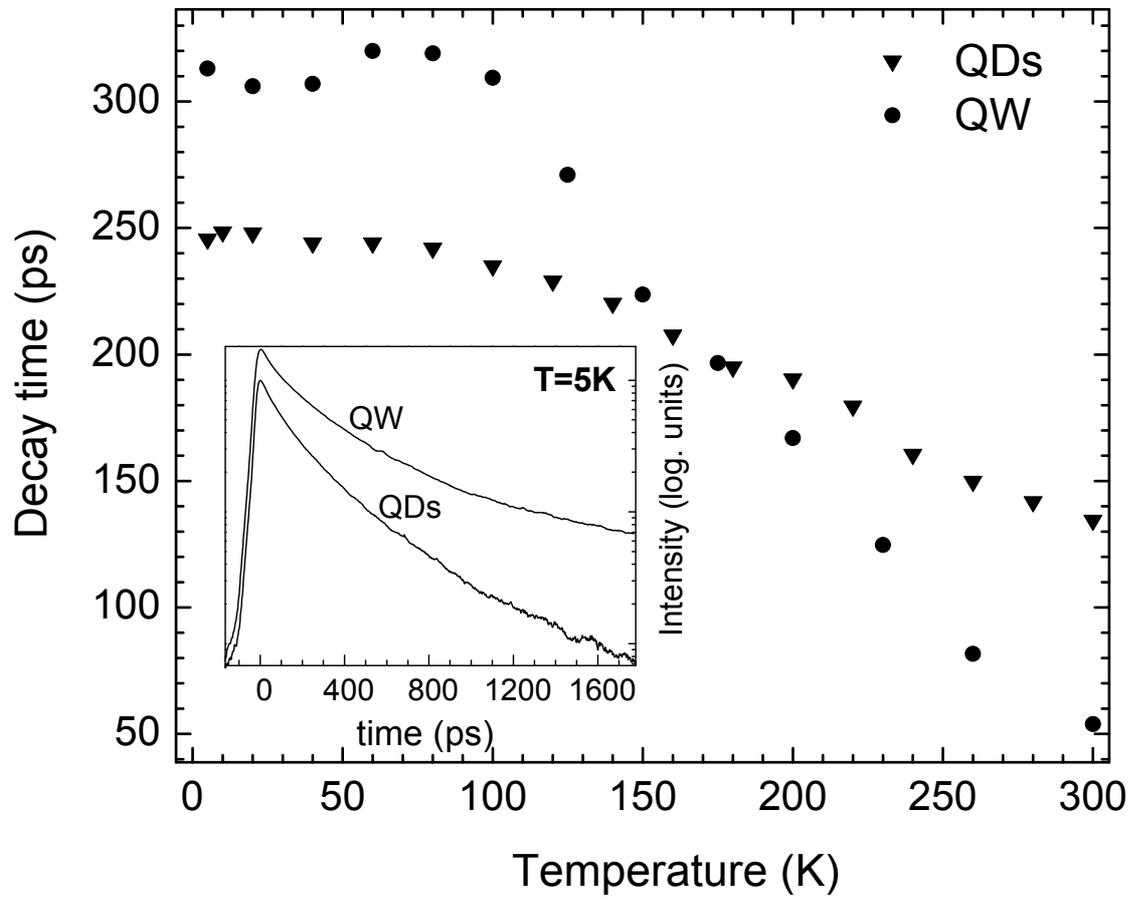

Figure captions :

Fig. 1 : Photoluminescence decay curves on ensembles of QDs at 4K for various detection wavelengths (a, b, c, d). The decay curves are not monoexponential and do not vary as a function of wavelength.

Fig. 2 : Time dependent spectral diffusion. The upper graph displays a single QD photoluminescence spectrum recorded as a function of time (100 consecutive spectra of 1 s integration time each). The graphs below show the variation of the emission energy compared to its time-averaged value $E_0$ and the evolution of the full width at half maximum of the line.

Fig. 3 : Temperature dependent microphotoluminescence on a single QD line.

Fig. 4 : QD lineshape analysis. The open dots display the same experimental data as Fig. 3. The dotted line is the Voigt function fit of the zero-phonon line, while the solid line is the result of the model described in the text that accounts for acoustic phonon emission and absorption. The inset in the upper right corner represents the solid line minus the dotted line, that is the contribution of the acoustic phonon emission and absorption to the emission line-shape at T=140 K.

Fig. 5 : Photoluminescence decay curves for three single QDs. The dashed lines are single exponential fits. The decays are clearly monoexponential and the decay times are not correlated to the emission wavelengths. The inset displays the time-integrated spectrum as obtained on the streak camera.

Fig. 6 : Evolution of the decay time for a-plane GaN/AlN QDs and an a-plane GaN/AlN QW (1.5 nm thick) as a function of temperature. The inset displays the decay curves at 5 K.